\documentclass{ws-p8-50x6-00}

\def\aa{{\em Astron. Astrophys.}}
\def\apj{{\em Astrophys. J.}}
\def\gamef{\gamma_{\rm eff}}
\def\gtsima{$\; \buildrel > \over \sim \;$}    
\def\gtrsim{\lower.5ex\hbox{\gtsima}}           
\def\ltsima{$\; \buildrel < \over \sim \;$}    
\def\lesssim{\lower.5ex\hbox{\ltsima}}           
\def\r{{\bf r}}
\def\u{{\bf u}}
\def\x{{\bf x}}

\def\JBP{J. Ba\-lles\-te\-ros-Pa\-re\-des}

\def\EVS{E. V\'azquez-Semadeni}
\def\VS{V\'azquez-Semadeni}

\begin{document}

\title{Turbulent Formation of Interstellar Structures and the
Connection Between Simulations and Observations}

\author{Enrique V\'azquez-Semadeni}

\address{Instituto de Astronom\' \i a, UNAM, Apdo.\ Postal 70-264,
M\'exico D.\ F.\ 04510, MEXICO \\E-mail: enro@astroscu.unam.mx}


\maketitle

\abstracts{
I review recent results derived from numerical simulations of the
turbulent interstellar medium (ISM), in particular concerning the
nature and formation of turbulent clouds, methods for comparing the
structure in simulations and observations, and the effects of projection of
three-dimensional structures onto two dimensions. Clouds formed as
turbulent density fluctuations are probably not confined by thermal
pressure, but rather their morphology may be determined by the large-scale
velocity field. Also, they may have shorter lifetimes than normally believed, 
as the large-scale turbulent modes have larger associated velocities
than the clouds' internal velocity dispersions. Structural characterization 
algorithms have started to distinguish the best fitting simulations to 
a particular observation, and have opened several new
questions, such as the nature of the observed line width-size relation
and of the relation between the structures seen in channel maps and
the true spatial distribution of the density and velocity fields.
The velocity field apparently dominates the morphology seen in intensity
channel maps, at least in cases when the
density field exhibits power spectra steep enough. Furthermore, the
selection of scattered fluid parcels along the line of sight (LOS) by 
their LOS-velocity inherent to the construction of spectroscopic data
may introduce spurious small-scale structure in 
high spectral resolution channel maps.}

\section{Introduction} \label{intro}

Turbulence is a prime example of a chaotic system, and the
interstellar medium (ISM) is most probably a prime example of a turbulent
medium.\cite{scalo87} A discussion of interstellar turbulence thus
befits this volume. Although chaos theory generally refers to systems 
with only a few degrees of freedom while turbulent flows have in general an
extremely large number of them, both types of systems share the
properties of sensitivity to initial 
conditions and the resulting practical unpredictability, as a
consequence of the nonlinear couplings between the relevant
variables. Furthermore, interstellar turbulence is much more complex
than natural terrestrial and laboratory turbulence because the former
is magnetized, and,  in cloudy regions, highly compressible and
strongly self-gravitating, thus not expected to Kolmogorov-like,\cite{VSPP4}
except possibly in the diffuse gas.
 
In recent years, many reviews covering various aspects of interstellar
and molecular cloud turbulence have appeared in the literature.
Compressible turbulence basics and self-similar models are discussed
by \VS.\cite{VS99} A compendium of a wide variety of
interstellar turbulence aspects is given in the volume {\it
Interstellar Turbulence},\cite{IT} including in particular turbulence
in the HI gas\cite{braun} and in the diffuse ionized
component.\cite{cordes,spangler,walterbos} A thorough review of the
implications of 
compressible MHD turbulence for molecular cloud and star formation is
presented in \VS\ et al.\cite{VSPP4} The present paper may be
regarded as a companion to the latter reference, as it includes a
number of topics not covered there. After reviewing the scenario of
interstellar clouds as turbulent density fluctuations and some of its
implications (sec.\ \ref{clouds}), I discuss the
comparison of the structural properties of simulated clouds with those
derived observationally for real clouds (sec.\ \ref{comparisons}; see
also Ossenkopf et al., this volume, for a 
discussion from a more observationally-oriented perspective), and
on the effects of projecting three-dimensional (3D) data onto two
dimensions, the latter point being important for the interpretation of
observational data, which necessarily are projections on the plane of
the sky (sec.\ \ref{projection}).

\section{Interstellar Clouds as Turbulent Density Fluctuations}
\label{clouds}

A fundamental problem in the understanding of star formation is how
the gas transits from a low-density diffuse medium to a comparatively
enormously denser star. An intermediate step in this process is the
formation of what we may generically refer to as interstellar clouds, including
structures that span a wide range of physical conditions, from large
diffuse HI clouds of densities a few $\times$ 10 cm$^{-3}$ and sizes up 
to hundreds of parsecs, to molecular 
cloud cores with densities $\gtrsim 10^4$ cm$^{-3}$ and sizes of a few
$\times$ 0.01 pc. 

The process of
cloud formation quite possibly involves more than a single mechanism,
including the passage of spiral density waves and the effects of
combined large-scale instabilities\cite{elm93a,elm95} operating
preferentially in the formation of the largest 
high-density structures, and the production of smaller density
condensations by either swept-up
shells,\cite{elmelm,vish83,vish94,hunter86} or by a generally
turbulent medium.\cite{hunter79,HF82,tohline87,elm93b} In the
remainder of this section we focus on the latter process.

An important question is whether structures formed by either turbulent
compressions or passages of single shock waves can become
gravitationally unstable and collapse. This depends crucially on the cooling
ability of the flow, which, as a first approximation, can be
parameterized by an effective polytropic exponent $\gamef$ such that
the pressure $P$ behaves as $P \propto \rho^{\gamef}$, where $\rho$ is 
the fluid density. Note that in this description both the ``cooling''
and the ``pressure'' can be generalized to refer to non-thermal energy
forms, such as magnetic and turbulent. 

The production and statistics of the density fluctuations in
polytropic turbulence has been investigated recently by various
groups.\cite{VS94,PVS98,NP99} Passot and \VS\cite{PVS98} have found
that the probability density function (PDF) of the density
fluctuations depends differently on the Mach number and on
$\gamef$. By means of a simple phenomenological model, these authors
find that, for isothermal flows ($\gamef=1$), the PDF is lognormal, as
a consequence of the Central Limit Theorem and of the accumulative and
multiplicative (additive in the log) nature of the density jumps
caused by shocks. Increasing the rms Mach number only increases the
width of the lognormal PDF and shifts its peak towards smaller
densities. Instead, varying $\gamef$ changes the form of the
distribution, which develops a power law tail at large densities for
$0 < \gamef <1$, and at small 
densities for $\gamef >1$. This effect is a consequence of the
modification of the 
lognormal by the local variation of the sound speed, which in the
general polytropic case varies with the density as $\rho^{(\gamef
-1)/2}$. Essentially, in the power-law side of the PDF the density
fluctuations are dominated by the nonlinear advection term in the momentum
equation, with an increasingly negligible contribution from the
pressure at increasingly large ($\gamef <1$) or small ($\gamef >1$) densities,
due to the decreasing sound speed, while on the opposite side of the
PDF the pressure dominates, impeding large excursions of the
density. A concise discussion of the mechanism can also be 
found in \VS\ and Passot.\cite{VSP99}

The stability of fluid parcels compressed in $n$ dimensions by shocks
or turbulence requires $\gamef > \gamma_{\rm cr} \equiv
2(1-1/n)$.\cite{mckee93,VPP96} For three-dimensional compressions,
the minimum Mach 
number $M_0$ necessary to induce collapse by the velocity field has
been computed by several authors as a function of $\gamef$ and the
mass $m$ of the cloud in units of the Jeans mass. It is found that
$M_0 \propto \ln m$ for the isothermal ($\gamef=1$)
case,\cite{hunter79} $M_0 \propto m^{(\gamef-1)/(4-3\gamef)}$ for $4/3 >
\gamef >1$(\cite{HF82}) and $M_0 \geq \sqrt{10/3(1-\gamef)}$ for
$0 < \gamef < 1$.\cite{tohline87}

A relevant implication is that clouds formed by turbulent
compressions are by necessity of a dynamical character, and are
expected to either collapse, if the conditions described in the
previous paragraph are satisfied, or else should reach some maximum
density and then ``rebound'' as the external compression subsides
(since the turbulent motions are chaotic, a compression will in
general last a finite time only). This result has interesting 
implications on two ``canonical'' concepts of interstellar
dynamics, as recently discussed by Ballesteros-Paredes et
al.\cite{BVS99} First, it may be that clouds in the ISM may not be
pressure-confined as in popular models of the
ISM,\cite{FGH69,wolfire} but rather in a highly dynamic and transient 
state, except if they become strongly gravitationally bound. An
interesting corollary of this scenario occurs for regimes in which the
{\it thermal} $\gamef \sim 0$ (i.e., a nearly isobaric behavior), as
is indeed the case for the ISM between densities $\sim 0.1$ and $10^2$ 
cm$^{-3}$.\cite{myers78} In this case, the thermal
pressure remains nearly constant, regardless of the density
structures formed by the turbulence. However, while traditionally this 
near pressure constancy has been regarded as a pressure balance
condition that provides confinement for clouds, in the turbulent-cloud 
scenario it is only a relatively irrelevant consequence of the medium
being maintained at nearly constant thermal pressure by the prevailing 
cooling processes as it is compressed by the turbulent motions.

Secondly,
it appears difficult to produce the quasi-hydrostatic clumps which are
the commonly assumed to be the initial conditions of many models of star
formation.\cite{shu87} Their formation within a globally
gravitationally {\it stable} region by a turbulent compression
requires, as described above, that $\gamef < \gamma_{\rm
cr}$. However, this sets them in a state of gravitational collapse. To 
then form a hydrostatic structure, a change in $\gamef$ is
required {\it during} the collapse, in order for $\gamef$ to now
become $\geq \gamma_{\rm cr}$, so that the pressure may now overcome the
ongoing gravitational compression, and ultimately halt the
collapse. Such change in $\gamef$, at least from thermal contributions 
alone, is not expected until very high densities ($\gtrsim 10^8$
cm$^{-3}$) are reached.\cite{BVS99}

Another implication of the turbulent density fluctuation scenario for the
clouds is that the time scales associated to clouds may be smaller than 
those derived from their sizes and internal velocity
dispersions.\cite{BHV99} If the cloud is made by the collision of
large-scale streams, the time scale for its formation
is the crossing time through its size $L$ at the velocity difference 
between the turbulent scales {\it larger} than cloud (the
colliding streams). For
turbulence with a (normal) spectrum that decays with wavenumber $k$, the
characteristic velocity difference increases with separation, and thus 
the crossing time scale for the cloud is smaller than that
derived from its internal velocity dispersion. Indeed, the line widths
from the HI envelopes of molecular clouds are generally larger than the
line widths in the molecular clouds themselves.\cite{BHV99} This result 
has been proposed as a possible solution to the absence of
post-T-Tauri stars in the Taurus region, since the region may be
younger than the age derived from its internal velocity
dispersion.\cite{BHV99}

\section{Comparisons Between Simulations and Observations}
\label{comparisons}

In recent years, numerical simulations of interstellar
turbulence have advanced to the point that statistical comparisons
with observational results have become possible.\footnote{But note that 
modeling of
individual objects is not feasible because of the sensitivity to
initial conditions of turbulent flows.} This is a crucial task because 
it allows an iterative procedure in which simulations may be
constrained as models of the ISM and interstellar clouds by comparison 
of their morphological, topological and statistical properties with
their observational counterparts. Once the best-fitting set of
parameters is found for a certain type of system, the simulations may
then be used as highly complete models of such system to improve our
understanding of the physical processes occurring within
them. However, it should be pointed out that this is a difficult task, 
because the natures of the observations and of the simulations are quite
different. While simulations are performed on a regular grid, with
well-defined boundaries, observations refer to regions of space for
which the size along the line of sight is not constant. For example,
for HI observations, the path length through the disk decreases with
Galactic latitude, while for molecular line observations, the observed 
objects in general may have different extents along the LOS over their 
projected area. Also, linear sizes perpendicular to the LOS increase with
distance. In spite of these difficulties, however,
several first steps have been taken in this direction.

\subsection{Scaling Relations} \label{scaling}

One basic question is whether the clouds in the simulations reproduce
the well-known Larson\cite{larson81} relations (see also the review by 
Blitz\cite{blitz93} for a more recent account) $\Delta v \propto
R^{1/2}$ and $\langle \rho \rangle \propto R^{-1}$, where $\Delta v$
is the velocity dispersion in the cloud, $\langle \rho \rangle$ its
mean density, and $R$ its characteristic size. In an analysis of
two-dimensional MHD simulations of the ISM including self-gravity and
stellar-like driving, \VS\ et al.\cite{VBR97} found that, although
with much scatter, a Larson-like velocity dispersion-to-size scaling
of the form $\Delta v \propto R^{0.4}$ is observed for clouds defined
as the connected regions in the flow with densities above a given
threshold. This result is roughly consistent with observational
surveys giving scaling exponents between 0.4 and
0.7.\cite{blitz93} However, the density-size relation is not verified in the
simulations. Instead, small clouds with low densities, which are
transient and not gravitationally bound, are formed in large
quantities in the simulations. Rather than being satisfied by all clouds,
the Larson density-size relation appears to be an upper bound to the
region populated by the clouds in a $\rho$-$R$ diagram. The same trend 
was observed in a sample of objects away from map intensity
maxima.\cite{falg92} This supports
suggestions\cite{kegel89,scalo90} that the density-size relation may
be an artifact of the limitations on integration times of
observational surveys, and that the $\Delta v$-$R$ relation, satisfied 
by all clouds, may
originate from the Burgers-like spectrum of molecular cloud
turbulence.\cite{VBR97} In this scenario, only those clouds which
become sufficiently self gravitating and are not strongly disturbed
(i.e., in near virial equilibrium), satisfy the density-size relation.

\subsection{Synthetic Line Profiles} \label{profiles}

Another important means of comparison between simulations and
observations are the spectral line profiles, since their shapes
(generally characterized by their first few moments) reflect the
velocity structure of the flow in the observed regions. Line profiles from the
simulations are constructed as density-weighted velocity histograms
along each LOS.

Falgarone et al.\cite{falg94} compared the line spectra produced in a
high-resolution 3D simulation\cite{PPW94} of weakly compressible
turbulence with observational data, concluding that both sets of
spectral lines are very similar, in terms of the range of values of
the variance and the kurtosis they present. The similarity is greater 
in this case than that achieved by other models of clouds (constructed 
with random uncorrelated velocity fields or with isolated clumps in an 
interclump medium) which do not
account for the spatial correlations derived from the continuum
nature of fluid turbulence. Similar results have been obtained
from randomly generated flows with an imposed Kolmogorov
spectrum.\cite{dubinsky95} However, a more recent study based on the
PDFs of the velocity centroids, rather than on the line
profiles,\cite{miesch99} suggests that in turn nearly incompressible
turbulence fails to capture some features of the centroid
PDFs of both molecular and HI regions, which exhibit a larger
degree of non-Gaussianity than those derived from incompressible or
weakly compressible turbulence.

Line profiles from simulations of strongly compressible MHD turbulence,
including radiative transfer, as well as other diagnostics, have
recently been compared to molecular 
line data by Padoan and coworkers to support their suggestion that
molecular clouds may actually be in a super-Alfv\'enic regime, rather
than a sub-Alfv\'enic one, as generally believed.\cite{arons75} Their
tests were based on two 3D simulations of MHD isothermal turbulence without
self-gravity, one super-Alfv\'enic, the other sub-Alfv\'enic. 
First, they noted that simulated line profiles from the
super-Alfv\'enic run seem to reproduce the observed growth of
line width with integrated antenna temperature\cite{heyer96} better
than those from the
sub-Alfv\'enic run.\cite{padoan98} Secondly, the super-Alfv\'enic run
seems to better match the observed trend of magnetic field strength
vs.\ gas density.\cite{PN99} Finally, the super-Alfv\'enic simulation
also more closely reproduces the observational
trend of the dispersion of extinction vs.\ mean extinction along
selected lines of sight.\cite{PN99} However, one caveat
remains before the super-Alfv\'enic model can be accepted: the
simulations lacked self-gravity, which could have pushed the
results of the sub-Alfv\'enic simulation closer to the observational
results, and the super-Alfv\'enic run away from them. Therefore,
similar experiments are needed with self-gravitating runs in order to
confirm this possibility.

\subsection{Higher-Order Statistics and Fractality Analyses} \label{stats}

The methods described in the previous section have taken into account
only the {\it velocity} information in the spectral data ``cubes''. We now
briefly discuss methods that have tried to take {\it spatial} information into
consideration as well. 

Spatial structure is often described by means of the autocorrelation
function (ACF), which measures the probability of finding equal values 
of a given physical variable at two different positions in space, as a
function of their separation.\footnote{However, it has been
pointed out by Scalo and Houlahan\cite{scalo84,HS90} that one limitation
of the ACF is that it cannot distinguish between hierarchically
nested or randomly distributed structure.} Early studies in this
direction measured the ACF of column density and of the line velocity
centroids, attempting to determine whether characteristic lengths
exist in the ISM\cite{scalo84,KD84,KD85,DK87}, with mixed
results. Recently, a variant of this approach termed the Spectral
Correlation Function (SCF) has been
introduced.\cite{roso99} The SCF measures the quadratic difference between
line spectra (on a channel-by-channel basis) at different positions on
a spectral-line map, in an attempt to include spectral as well as
spatial information in the statistical description. So far, the method 
has been used to measure the angle-averaged correlation between the
spectrum at a given position in a map and at its nearest neighbors,
allowing the characterization of the small-scale variability of the
spectra, and a comparison between CO maps and simulations of
isothermal turbulence under various regimes (purely hydrodynamic, MHD, 
and self-gravitating). In that work, differences between the values of the
SCF for weakly compressible purely hydrodynamic turbulence\cite{PPW94} 
and for the Ursa Major molecular cloud indicate that simulation is not 
as accurate a model for the Ursa Major cloud as previously claimed by
Falgarone et al.\cite{falg94} on the basis of line profile shapes (see
sec.\ \ref{profiles}). Comparison of HI data with non-isothermal
simulations has proven more difficult, because the thermal broadening
of the warm gas swamps the velocity structure of the cold gas.\cite{BVG99}

The recognition that the structure of the ISM may have a turbulent
origin has also prompted searches for fractal properties of
interstellar clouds (although see the contribution by Combes, this
volume, for an alternative scenario originating the fractal
structure). Early studies in this 
direction\cite{scalo90,falg91} started by 
measuring the fractal dimension of the clouds in column density or
intensity maps of selected regions, by means of the
area-perimeter scaling in the projected clouds, finding dimensions
near 1.4. Recent measurements of their area-perimeter relation in 2D
simulations find similar values.\cite{BP99} However, these values 
are surprisingly close to the projected fractal
dimensions of clouds in the Earth's atmosphere\cite{love82},
suggesting that the fractal dimension may be a significantly
degenerate diagnostic which may not be capable of distinguishing
between different physical regimes. Indeed, a more recent  
study by Chappell and Scalo\cite{chapp99} has shown that the column
density maps of various regions actually have a well-defined
multifractal structure, so that attempting to measure a single fractal
dimension for the clouds in the maps may erase much of the structural
information. Furthermore, these authors emphasize that the methods
used to determine fractal dimensions of clouds rely on the definition
of ``clouds'' by means of some rather arbitrary criterion (such as
thresholding the column density field), while the multifractal
spectrum determination uses the structural information of the whole
field. The multifractal spectrum of the regions studied
appears to correlate fairly well with the geometric forms seen
visually, potentially providing a means for quantitative structure
classification schemes. The multifractal properties of numerical
simulations of ISM turbulence are currently being
investigated.\cite{chicana} 

A method for determining the line width-size scaling of spectral maps
independently of any specific definition of ``clouds'' in a spectral
map has been introduced recently by Heyer and
collaborators.\cite{HS97,HB99} The method uses the statistical
technique known as Principal Component Analysis (PCA) to define a set
of spectral profiles ({\it eigenvectors}) which form a ``natural''
basis for the spectral 
maps (in the sense that it reflects the main trends of the intensity
data among the velocity channels). {\it Eigenimages}, which represent
the intensity structures as ``filtered'' by the basis spectra, are
generated by projecting the original spectra onto 
the eigenvectors. By then measuring the ACF of the eigenvectors and
of the eigenimages, the relationship between the magnitude of velocity
differences and the spatial scales over which these differences occur
can be extracted. In order to ``calibrate'' the
expected value of the exponent $\alpha$ in this relation (such that
$\Delta v \propto R^\alpha$, where $\Delta v$ is the velocity difference and
$R$ is the size), ``pseudo-simulations'' of fractional Brownian noise
with a prescribed spectrum were produced and ``observed'' using a
radiative transfer simulator.\cite{HB99} It was found that 
the scaling exponent is related to the spectral index $\beta$ by
$\alpha =\beta/3$. Recent calibrations with actual hydrodynamic and
MHD\cite{HBPV99} simulations in 3D are roughly consistent with this
result, although the resolutions available in 3D simulations are still 
insufficient to develop clear power-law turbulent spectra, so the
results are not conclusive. Tests with higher resolution 2D
simulations apparently produce a different calibration,
$\alpha=\beta/4$. 

The origin of these scalings is not well understood
yet. In fact, they suggest that the very nature of the line width-size
relation derived through PCA is unknown, since it does not follow
the same scaling as, say, the second-order structure function, defined 
as $F_2(r)=\langle [\u(\x)-\u(\x+\r)]^2 \rangle$, where the brackets
denote a volume average. This function gives
the mean quadratic velocity difference between positions separated a
distance $r$. Yet, $F_2$ is related to the energy spectrum by\cite{lesieur}
\begin{equation}
F_2(r)=4 \int_0^\infty E(k) \bigl(1-\frac{\sin kr}{kr}\bigr) dr,
\end{equation}
so that, if $E(k) \propto k^{-\beta}$, then $F_2(r)\propto r^\eta$, with
$\eta= (\beta-1)/2$. $\eta$ is thus not functionally related to
$\beta$ in the same manner as $\alpha$. The discrepancy is probably
related to the fact that the above scaling refers to the structure function 
of the actual 3D velocity field, while in spectroscopic data every
velocity interval contains the contribution of many fluid
parcels (possibly far apart from each other), and only one of the
three velocity components is observed. Thus, the true nature
of the spatial and velocity increments in spectral data cubes 
remains unknown.

Finally, a method for structure analysis similar to the power
spectrum, but using a wavelet rather than a Fourier basis is described 
in the contribution by Ossenkopf et al.\ (this volume).

\section{Effects of Projection on Morphology} \label{projection}

One advantage of 3D simulations is that they contain more structural
information than even spectroscopic data ``cubes''. While the latter 
only refer to two spatial and one velocity dimensions, 3D numerical
simulations provide information on the 3D structure of all physical
variables. (Of course, their downside is that they are necessarily
limited in resolution and in the number of physical processes that can 
be included.) This allows an investigation of the 3D structures that
generate the patterns observed in the position-position-velocity
(PPV) space of the spectroscopic channel maps. To this end, channel
maps are constructed from the simulations by integrating the density
field along one direction (the line of sight, or LOS), and then
selecting the contribution of each parcel along the LOS by its
LOS-velocity. This is equivalent to constructing density-weighted
velocity histograms (the ``line spectra'') at each position in the
plane perpendicular to the LOS. 

A rather unexpected result has recently been found independently,
using different approaches, by several groups. It appears that,
at least under certain conditions, the projected {\it spatial
structure in the channel maps} is dominated by the spatial distribution
of the {\it velocity} field, rather than by the 3D density
field. Pichardo et al.\cite{pichardo99} have shown this in a 3D
simulation of the ISM at intermediate scales (3--300 pc) by noting
that the pixel-to-pixel correlation between channel maps and thin slices of
the 3D velocity field tends to be larger on average than the
correlation between channel maps and slices of the 3D density
field. Independently, Lazarian and Pogosyan\cite{LP99} have shown
analytically that, for cases with an underlying one-to-one mapping
between the position along the LOS and the LOS-velocity (as for an
expanding universe or the HI gas distribution in the Galaxy), and with 
uncorrelated random density and velocity fields with well-defined
spectral indices, 
the power spectrum of the projected density field is dominated by the
spectrum of the velocity field for density spectral indices steeper
than $-3$, unless the velocity channels are very wide (as is clearly
the case in the limit of a single velocity channel, in which the
velocity dependence is integrated out). Finally, Heyer and
Brunt\cite{brunt99,HB99} have noticed that, in their
pseudo-simulations (cf.\ sec.\ \ref{stats} above), channel maps and
PCA-derived $\Delta v$-$R$ relations produced
with and without density-weighting are notoriously similar, suggesting 
that the effect of the density weighting is relatively minor. A
similar effect was noticed by Falgarone et al.\cite{falg94} about the
shapes of synthetic line profiles. In summary, it appears that the
spatial structure of the velocity field is at least as important as
that of the density field in determining what is observed in projection 
on the plane of the sky.

A related effect has been observed by Pichardo et
al.\cite{pichardo99} The morphology observed in the channel
maps appears to contain much more small-scale structure than either
the density or the velocity 3D fields. This is reflected in
the power spectra of the channel maps and of 2D slices through the 3D
density and velocity fields, the latter two having steeper slopes and
falling off much more rapidly than the former. This phenomenon has
been interpreted by those authors as a consequence of a pseudo-random
sampling of fluid parcels along the LOS by the velocity selection
performed when constructing a channel map. This introduces an
additional ingredient of variability between neighboring LOSs, which
causes artificial small-scale variability in the channel maps.

It can be concluded from this section that, for a fully turbulent ISM, 
the structure seen observationally, through spectroscopic
observations, may differ from the actual 3D structures
present in the medium. In particular, this suggests that
structure-finding algorithms operating on spectroscopic data cubes may 
not identify exactly the same structures than would be obtained on the
actual 3D spatial data, as already pointed out by various
authors.\cite{AR92,brunt99} These effects may be decreased, however, in cases 
when the observed regions contain well-defined ``objects'' which may
be picked out by the observing process, such as shells, bipolar flows, 
etc. 

\section{Conclusions} \label{conclusions}

In this review I have discussed recent results from numerical
simulations of turbulence in the ISM. I first reviewed the scenario of
interstellar clouds as turbulent density fluctuations. Work on
the production of gravitationally bound structures in globally stable
media by turbulent compressions was summarized, in particular the
necessary Mach numbers\cite{tohline87} (for 3D compressions) and the
constraints on the effective polytropic exponent $\gamef$ for $n$-dimensional
compressions.\cite{VPP96} Three implications were then
discussed.\cite{BVS99,BHV99} First, 
the near thermal pressure balance observed in the ISM (except for
molecular clouds) may not be a confining agent for clouds, but rather
a relatively fortuitous consequence of the prevailing heating and
cooling mechanisms, which render the medium nearly isobaric, in the
presence of turbulence-induced density fluctuations. Second, it
appears unlikely that nearly hydrostatic cores may be produced within
the turbulent ISM unless some very specific variations in $\gamef$
occur during a gravitational contraction induced by the
turbulence. Third, the time scales associated with clouds may be
smaller than those estimated from the clouds' characteristic
dimensions and their velocity dispersions, since the relevant
velocities may instead be those of the larger, external flow streams that
produced the clouds at their collision interfaces.

I then proceeded to review recent results from various attempts to
relate numerical simulations to observational
data, from early qualitative comparisons of spectral line
profiles\cite{falg94} and surveys of clouds in 2D simulations (which
suggested the existence of a whole population of low-column density
clouds that do not satisfy Larson's density-size
relation)\cite{VBR97}, to recent approaches using more sophisticated
statistical techniques such as the Spectral Correlation
Function\cite{roso99}, Principal Component Analysis\cite{HS97,HB99},
and fractal dimensions and multifractal
spectra\cite{scalo90,falg91,chapp99}, mostly aiming at characterizing
the morphology of interstellar structures in a statistically
meaningful way, and determining whether the structures developing in
turbulence simulations reproduce their properties. Some of these
methods are only being developed now, but they are already providing a
quantitative method for discriminating between turbulence simulations
with different parameter choices as the most suitable models for
specific interstellar regimes, as well as providing a basis for
interpreting observational data in terms of the simulations.

Finally, I discussed the relationship between the actual 3D spatial
structure of the density and velocity fields, and that of the
projected ``intensity'' field in channel maps. Recent works, using
various approaches\cite{HB99,pichardo99,LP99}, suggest that the
structure, both morphological and
statistical (power spectrum) of the 2D intensity field is dominated by 
the velocity field, rather than by the density. Additionally, since
the total intensity in every LOS of a channel map is constructed by
``selecting'' scattered fluid parcels (on the basis of their LOS velocity) 
along the LOS, spurious small-scale variability is introduced into the
structure seen in the 
channel maps, as the set of sampled parcels varies from one LOS to the 
next.\cite{pichardo99} These results suggest that the structure in
channel maps bears a  
very complex and non-trivial relationship to the structures actually
existing in the ISM. Further work in this area is likely to produce
numerous unexpected and exciting results in the near future.


\section*{Acknowledgments}

I am grateful to acknowledge Chris Brunt, 
Mark Heyer, Alex Lazarian, Volker Ossenkopf, Dimitri Pogosyan and John
Scalo for stimulating discussions and useful comments. This work has 
received financial support from grants CRAY/UNAM SC-008397 and
UNAM/DGAPA IN119998.

\end{document}